\documentclass[aip,rsi,reprint]{revtex4-1} 
\usepackage{xcolor,graphicx}
\usepackage[utf8]{inputenc}
\usepackage{url}
\usepackage{epstopdf}
\usepackage{textcomp}
\usepackage{color}

\begin{document}

\title{An autonomous robot for continuous tracking of millimetric-sized walkers} 

\hyphenation{experiments resolution camera}

\author{A. Serrano-Muñoz}
\affiliation{Group of Complex Systems and Statistical Physics, Physics Faculty, University of Havana, 10400 Havana, Cuba}
\author{S. Frayle-Pérez}
\affiliation{Group of Complex Systems and Statistical Physics, Physics Faculty, University of Havana, 10400 Havana, Cuba}
\affiliation{DSIC, Technical University of Valencia, 46022 Valencia, Spain}
\author{A. Reyes}
\affiliation{Group of Complex Systems and Statistical Physics, Physics Faculty, University of Havana, 10400 Havana, Cuba}
\author{Y. Almeida}
\affiliation{Faculty of Mathematics and Computer Science, University of Havana, 10400 Havana, Cuba}
\author{E. Altshuler}
\author{G. Viera-López}
\email[]{gvieralopez@fisica.uh.cu}
\affiliation{Group of Complex Systems and Statistical Physics, Physics Faculty, University of Havana, 10400 Havana, Cuba}

\date{\today}

\begin{abstract}
\rightskip.5in
The precise and continuous tracking of millimetric-sized walkers --such as ants-- is quite important in behavioral studies. However, due to technical limitations, most studies concentrate on trajectories within areas no more than 100 times bigger than the size of the walker or longer trajectories at the expense of either accuracy or continuity. Our work describes a scientific instrument designed to push the boundaries of precise and continuous motion analysis up to 1000 body lengths or more. It consists of a mobile robotic platform that uses Digital Image Processing techniques to track the targets in real time by calculating their spatial position. During the experiments, all the images are stored, and afterwards processed to estimate with  higher precision the path traced by the walkers. Some preliminary results achieved using the proposed tracking system are presented.

\end{abstract}

\maketitle 

\section{INTRODUCTION}

Animal tracking has been a continuous challenge for scientists and engineers alike. Several approaches have been used in this field of research such as RFID, SODAR, SONAR, X-Rays, Computational Tomography and Computer Vision \cite{remote-sensing}. 

Fixed Camera tracking (FC) is the most commonly used approach to record and analyze the motion of arthropods, and other species or moving objects \cite{multi-camera-tracking}. In FC, one or more image acquisition devices are statically deployed at a given height, covering the surface to be studied. This surface is, of course, limited by the camera's field of view. The experimental setup is quite simple and results are not affected by vibrations or displacements, given that the image acquisition device remains in the same position. However, the major drawback is the limited observation area and duration of the experiments. Research efforts have been carried out using this technique to study not only single individuals, but collective motion as well \cite{live-insect-colonies,multiple-ant-tracking}. The region of study could be further increased, by raising the height of the cameras, but in doing so image resolution and therefore precision in the estimated positions may suffer substantially \cite{reyes2016preliminary}. 

Following another approach, A. Narendra \emph{et al.} mounted a differential GPS on a support being held by one human operator throughout their experiment \cite{narendra2013mapping}. This mobile approach allows one to obtain longer trajectories but the resolution of the data is low since the GPS uncertainty can be as big as 1.0\,cm which easily exceeds the size of the insect body. The researchers must chase the insect, while carrying the load of the instrument. This is indeed a very labor intensive task, especially in hot environments.

The work of H. Dahmen \emph{et al.} \cite{treadmill2017} stands out as an example of mobile region procedures applied to insect tracking. The instrument consists of an air-cushioned lightweight spherical treadmill that registers the path of an animal walking on top of the sphere. Long trajectories are obtained with a high degree of precision. However, this system may be invasive to the insect, potentially altering its natural behavior.

In this work we study the motion of single individuals in non confined areas. We have designed and implemented a system able to track millimetric-sized walkers --such as many species of insects and even crustaceans-- moving over a few meters on a flat surface. This opens new possibilities in the field of behavioral ecology.

Scientific instruments built using automated computer control and robots are becoming more widespread \cite{cornell2008automating,gibouin2018study, baniqued2018biomimetics}. In this work we use a differential drive robot to keep track of the walker. In principle, other robots able to maintain a localization of their position and equipped with a camera could be used with our techniques, such as flying robots or other kind of wheeled mobile robots \cite{VENTURA201685,6696922,7429411}.

This paper is organized as follows: in Section \ref{sec_instrument} the tracking instrument is fully described. Section \ref{sec_procedure} presents the procedure of a typical experiment based on the proposed apparatus. In Section \ref{sec_measuerd_variables} the variables being measured are described. Section \ref{sec_results} includes the evaluation of the uncertainty based on the tracking of artificial walkers, and also the tracking of real individuals belonging to two different species of arthropods.


\section{The Tracking Instrument}
\label{sec_instrument}

Current methods of tracking walkers have limitations in accuracy or tracking length. The complexity and hard work required to track walkers in areas over 1000 times larger than the size of the individual --``non-confined'' areas--\cite{narendra2013mapping} have led us to design an instrument that accomplishes this task in a fully automated way. Our instrument exploits the accuracy of the fixed camera approaches, but changes the position of the camera based on the animal’s location to explore bigger areas.

\begin{figure}[!ht]
    \begin{center}
        \includegraphics[width=280px]{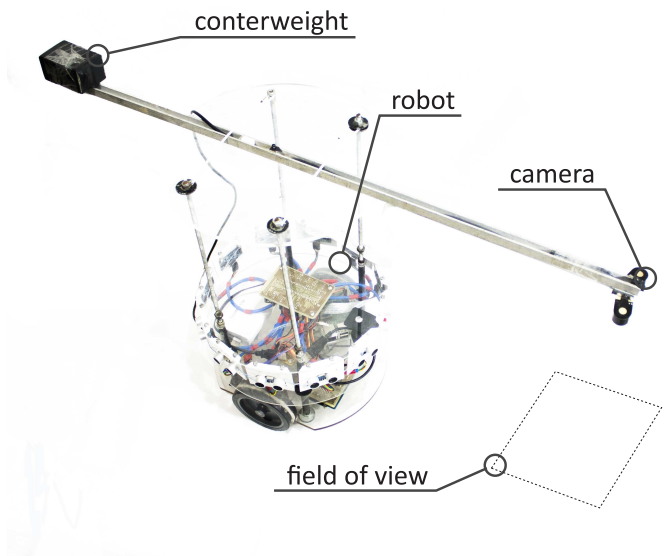}
        \caption{Sketch of the instrument based on a Differential Drive mobile robot used to track non-flying arthropods. The infrared camera on top, attached at the end of the horizontal arm, allows to work on low illumination conditions. The outer diameter of the mobile robot is 28\,cm and the bar supporting the camera is 80\,cm long.}
        \label{fig-experimental-system}
    \end{center}
\end{figure}

The instrument is a differential drive mobile robot able to track millimetric-sized walkers for long time intervals without human interaction. We chose this particular kind of robot because it is easy to localize it using the motion of its wheels. But with minor changes, other kinds of robots can be used depending on the desired working environment.  In Figure \ref{fig-experimental-system} a sketch of the instrument and all its components is shown. 

An infrared camera is the primary sensor used to detect the position of the walker. It was placed on the robot to capture images of the ground covering an area of $0.1054\,m^2$ at approximately 30\,cm in front of the robot. The camera type can be varied, but is important that the resolution of the camera be high enough to resolve the walker. Using an infrared camera allows the instrument to work at night which is convenient for the study of nocturnal species.  The images are scaled via GPU at $512\times512$ pixels and processed to find the position of the targeted individual in each frame. The decision of whether it is necessary to move the robot in order to keep the target in the field of view of the camera is made based on the position of the walker relative to the camera.

The robot was designed using open source technologies. Motors with 360 steps-per-revolution Hall effect encoders and a custom motor driver based on Arduino were chosen. The core of the system is the single board computer Raspberry Pi model 3. It handles the communications via WiFi or Ethernet, motion control and localization of the robot, and the camera sampling as well as logging and processing of images. However, in most cases an external computer is used to process the images in real time in order to obtain a lower latency in the determination of further movements.

The temporal evolution of the position of the robot is stored along with the images acquired during the experiments. By processing these images it is possible to locate the walker relative to the camera. Using the information of the position of the robot in the map, it is then possible to locate the walker relative to the area.

\section{Tracking Procedure}
\label{sec_procedure}

In our previous work\cite{chasing-insects}, several tracking algorithms based on image processing designed to capture the trajectory of a single insect in a sequence of frames were presented and discussed, focusing on the ones involving a mobile camera. The instrument proposed  here is able to work with all of these algorithms. In this section we explain the procedure of a typical experiment, covering the details of every step of the process.

\subsection{Work-flow}
\label{sub:workflow}

At first, the robotic platform is carried near the target individual, placing the camera right above it. At this point the trail left by the walker begins to be traced. The robot, and therefore the camera, remain static, hence the Frame Differencing algorithm\cite{chasing-insects} may be applied, although other algorithms can be used. When the target moves, altering its current location, it will be detected and a Region of Interest (ROI) around it will be selected. 

As the walkers explores the area, it will eventually escape the field of view of the camera. The robot must act right before this happens, rapidly moving in such a way that the target occupies again a spot near the center of the visual field of the camera, as it did at the beginning of the experiment. To be able to determine when to move, it is necessary to track the walker in real time. This Real-Time Tracking does not need to be extremely accurate, but it needs to work as fast as possible. 

Once the data from the whole experiment is gathered, another tracking process is performed in order to obtain an accurate estimation of the position of the walker relative to the camera. The trajectory of the individual is reconstructed relative to the area using the images and the position of the robot.

\subsection{Real-Time Tracking}
\label{sub:RTTracking}

\begin{figure}[!b]
        \includegraphics[width=180px]{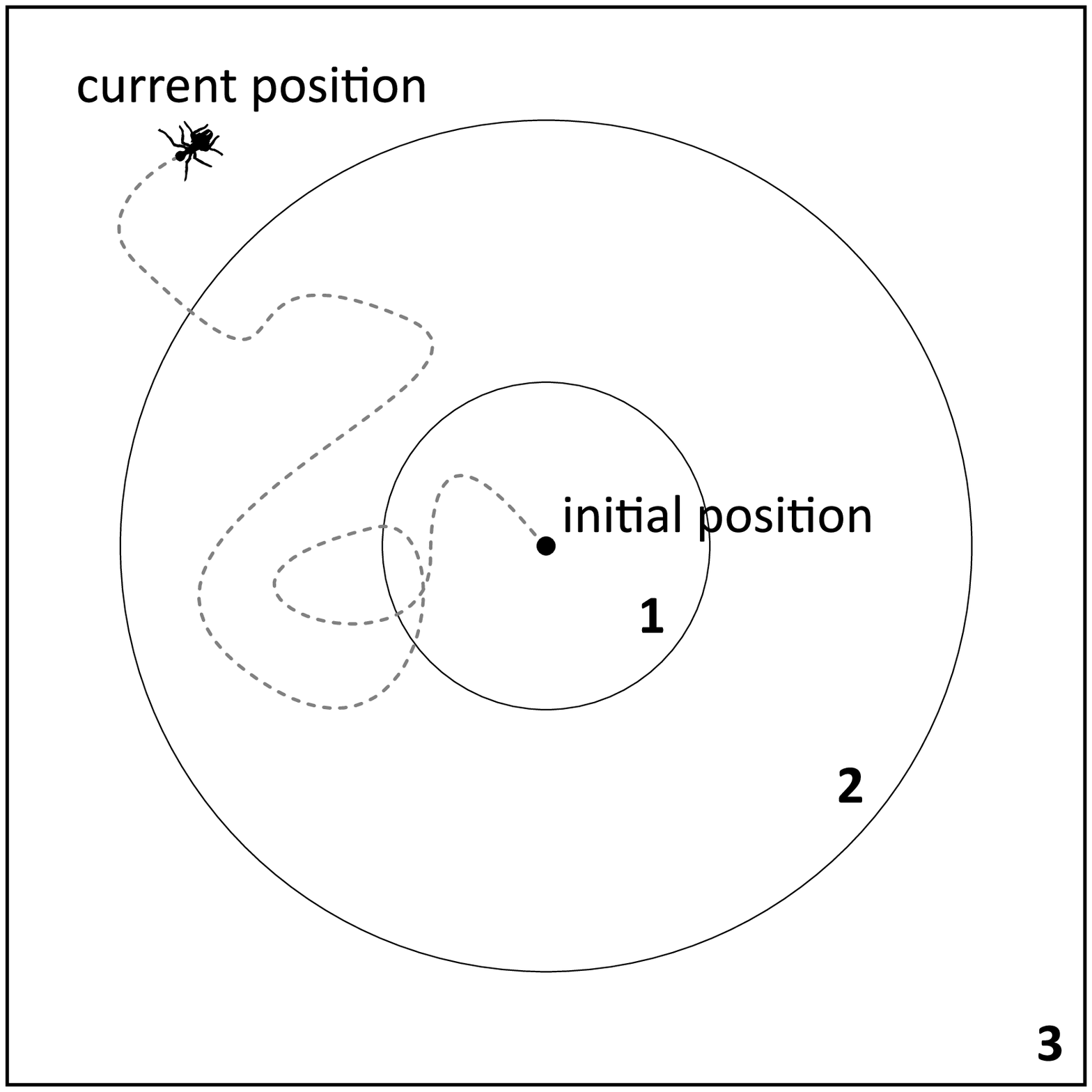}
        \caption{Classification of frames in regions labeled 1, 2 and 3 based on the possibility of the target to escape from the camera's field of view in the following frames.}
        \label{fig_zones}
\end{figure}

High accuracy in the real-time tracking is not important, because all of the camera frames captured and robot positions are stored for further processing in order to obtain a highly accurate trajectory. The goal at this stage is to keep the target walker inside the field of view of the camera at all times. To accomplish this task, each frame has been divided into three regions, labeled as 1, 2 and 3 in Figure \ref{fig_zones}. The radii of the circumferences enclosing regions 1 and 2 can be easily modified, since not all walkers move at the same speed.

The outermost region is used as a trigger zone, to indicate that the walker is near the edge of the field of view. The robot must then move in such a way that the target is centered. It will keep moving, thereby transitioning the target from the region labeled 3, to the one labeled 2 and finally 1. Once it has been positioned in this inner zone, the robot stops, until a new escaping threat is detected, and the procedure then repeats. 

\subsection{Trajectory Reconstruction}
\label{sub:trajrec}

At this point, the purpose is to obtain the actual trajectory of the individual being tracked with high precision, regardless of the processing time. This is better done after gathering the data in the field, in order to avoid excessive CPU usage that may slow down the data acquisition process or allow the target to escape the field of view of the camera.

The tracking process of the individual in this stage is highly accurate, but may require human assistance when the tracked object cannot be auto-detected accurately. All the images gathered are reprocessed to acquire the position of the target arthropod relative to the camera on every frame. Next, we write the position of the arthropod relative to the area in which the origin is the arthropod's initial position in the ground. The problem of reconstructing the trajectory of the walker on the area can be solved if the trajectory of the camera relative to the area can be obtained.

\subsection{Camera Positioning}
\label{sub:campos}

As the camera is located at a fixed angular and lineal distance relative to the robot, both will move together. So, localizing one of them, will automatically provide the position of the other. The localization of the camera can be achieved by several methods\cite{kowadlo2008robot, fuentes2015visual, dellaert1999using, yassin2016recent}. We implemented two different methods to localize the camera that are useful under different circumstances.

\subsubsection{Robot Odometry}
\label{subsub:ro}

The first localization method aims to estimate the camera's position via Robot Odometry \textbf{(RO)}. This procedure integrates the rotation of the wheels in order to compute a relative location of the robot following the equations: 
\begin{equation}
\Delta S  = \frac{r}{2}  (\Delta\phi_R + \Delta\phi_L)
\label{eq:deltaS}
\end{equation}
\begin{equation}
\Delta \Theta  = \frac{r}{d}  (\Delta\phi_R + \Delta\phi_L)
\label{eq:deltaTheta}
\end{equation}
where $ \Delta S $ is the linear displacement of the robot, $ \Delta \Theta  $  is the angular displacement of the robot, $ r $ is the radius of the wheels, $ d $ is the distance between the wheels and $ \Delta\phi_R $ and $ \Delta\phi_L $  are the angular displacement of the right and left wheels respectively.

Based on the incremental magnitudes and the initial coordinates of the robot relative to the area it is possible to estimate its current localization $(x, y, \Theta)$ for each time step $k$ in coordinate system centered in the initial position of the robot. The localization can be obtained using a second order Runge-Kutta integration through the following equations:

\begin{equation}
X_k  = X_{k-1} + \Delta S \cos(\Theta_{k-1} + \frac{\Delta\Theta}{2}) 
\label{eq:rungeX}
\end{equation}

\begin{equation}
Y_k  = Y_{k-1} + \Delta S  \sin(\Theta_{k-1} + \frac{\Delta\Theta}{2}) 
\label{eq:rungeY}
\end{equation}

\begin{equation}
\Theta_k  = \Theta_{k-1} + \Delta \Theta
\label{eq:rungeZ}
\end{equation}

Once the robot is localized, the camera position is known through Equation \ref{eq-camera-robot-odometry}, where w is the camera’s distance from the center of the robot:

\begin{equation}
    \left(\begin{array}{c} X_c \\ Y_c \\ \Theta_c \end{array}\right) = w \left(\begin{array}{c} \cos \;\Theta_r \\ \sin \;\Theta_r \\ 0 \end{array}\right) + \left(\begin{array}{c} X_r \\ Y_r \\ \Theta_r \end{array}\right)
    \label{eq-camera-robot-odometry}
\end{equation}

This method is an effective way of estimating the localization of the camera in environments with uniform floors where the wheels are not likely to slip. The computed localization is independent of the illumination of the environment.  But, as any relative localization method, it also carries an accumulative error due to the numerical integration.

\subsubsection{Monocular Visual Odometry}
\label{subsub:mvo}

The other approach used was the Monocular Visual Odometry \textbf{(MVO)}. This method directly estimates the position of the camera based on the detection of features in the captured images \cite{mvo}. The procedure extracts features from a first image and tries to find the same features in a second image. Afterwards, a linear transformation matrix (rotation, scale and translation) is calculated for these features. Finally, the position of the camera is estimated based on the calculated transformation matrix by a numerical integration of the relative displacements on each frame. 

The scale value of the calculated matrix can be used to evaluate the quality of the computed transformation, because the scale has to be constant as the camera is not changing its height relative to the ground. In case the scale changes significantly, new features have to be acquired.

This method tracks the position of the camera directly and does not depend on the amount of slippage of the wheels. However, it requires that the floor has enough landmarks to be used as features for the algorithm; it still carries a cumulative error due to the numerical integration process.

Generally in our experiments, we use MVO unless the floor is too homogeneous because it has fewer sources of error. However, other methods of localization can be tested to increase the accuracy in this stage.


\section{Measured Variables}
\label{sec_measuerd_variables}

Different parameters can be used to quantify the motion of animals during free exploration, some of them have been used in very diverse fields. These parameters characterize the walker trajectory when interacting with its surroundings. For example, \emph{Escherichia coli} bacteria can be tracked in a three dimensional liquid solution medium \cite{berg1972,darnige2017lagrangian}, and insects such as ants in a bi-dimensional medium \cite{bidimensional-ant-environment}. 


Two of the parameters useful to characterize the trajectory of walkers are the diffusivity and turn symmetries \cite{alfredo2016}. Both are briefly described next. 




\subsection{Diffusivity}
\label{sub:diffusivity}

Diffusivity relates the average change in position of a particle or set of particles in a given time using their Mean Squared Displacement (MSD) $\equiv <r^2>$, where r is the distance moved and the brackets refer to an ensemble average. Normal diffusion (or Brownian motion) follows an $<r^2> \sim t$ law\cite{majumdar2010universal,viswanathan2011physics}. More generally, we can define the law $<r^2> \sim t^\gamma$, where the slope $\gamma$ is the parameter used to define the diffusivity. The MSD is calculated as the difference between the present position and the initial position for each instant of time. Following the classification of Viswanathan \emph{et al.}\cite{viswanathan2011physics}, the different regimes of motion can be classified, based on the value of $\gamma$, as super-ballistic, ballistic, super-diffusive, normal diffusive and sub-diffusive.

\subsection{Turn Symmetry}
\label{sub:turnsymmetry}

Turn Symmetry is used to characterize the angular changes in a trajectory. It finds turning patterns by means of a rotation histogram\cite{viswanathan2011physics}. To generate the data for the histograms, it is necessary to compute all the angular positions of the walker relative to a fixed coordinate system. Consecutive angles are subtracted to obtain relative rotations at each sample. Finally, the relative angles are ordered, forming a rotation histogram that shows how likely is the walker to rotate at a given angle.

In order to analyze these variables correctly, it is necessary to obtain for all time steps the position of the individual with a high precision in a sufficiently long lasting experiment. Useful trajectories should be, at least, three orders of magnitude larger than the size of the individual.

\section{Results and Discussion}
\label{sec_results}

Validating this kind of system is a challenging task. When using individuals of the same species, or even the exact same individual, it is possible to get different results when quantifying the variables of interest, due to fluctuations in the animal behavior. Getting the ground truth of the walker’s actual trajectory is not easy. Next, we present different results in the tracking of artificial and real walkers.

\subsection{Tracking of virtual walkers}
\label{sub_restrackvirtual}

In order to validate our tracking system, we designed an experiment in which a virtual walker is generated and projected on a LCD screen placed face-up on the ground. We then had the robot track the walker. This configuration allows to compare a precisely generated trajectory of a walker with the tracked trajectory obtained using our system. Even in the limited region of the LCD screen, it is possible to generate long trajectories that make the walker move 1000 times its size, causing the robot to move several times, thus accumulating error in the localization.

\begin{figure}[!ht]
    \begin{center}
        \includegraphics[width=200px]{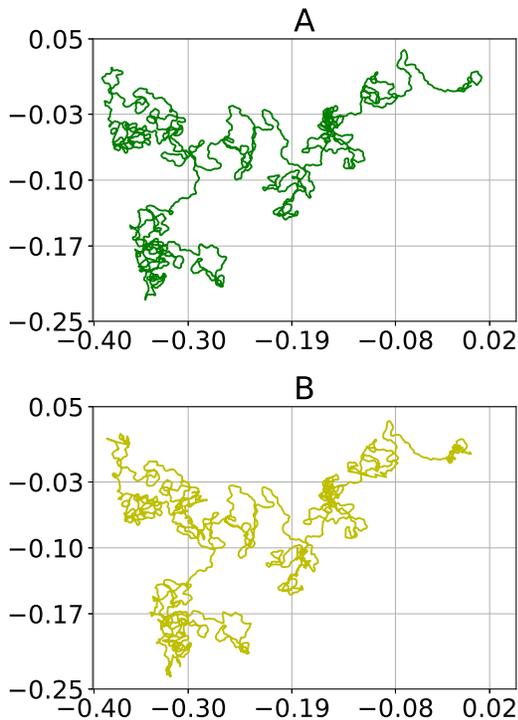}
        \caption{Tracking a virtual bug. (A) generated trajectory of a virtual walker that was projected on a LCD screen. (B) trajectory produced by our instrument, after tracking the virtual walker shown above.}
        \label{pathcomp}
    \end{center}
\end{figure}

The virtual walker was programmed to rotate a random angle on each time step following a Gaussian distribution. If the walker was not at the borders of the screen, it would move forward a constant distance. That simple automaton produces very complex trajectories. Figure \ref{pathcomp}A shows a sample generated trajectory, of around 7\,m long, that was used to validate the system. The robot was placed at different initial positions and then tracked the virtual walker 10 times. In Figure \ref{pathcomp}B it is possible to observe a reconstructed trajectory obtained as a result of one of the tracking processes performed.

\begin{figure}[!ht]
    \begin{center}
        \includegraphics[width=240px]{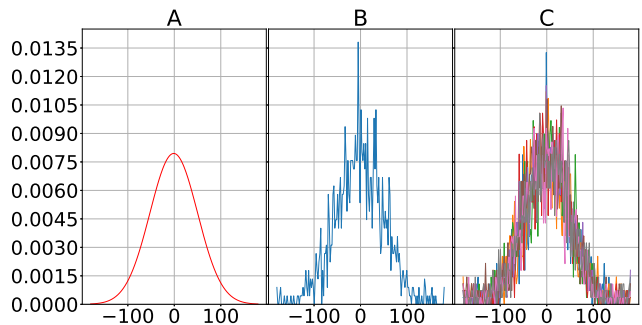}
        \caption{Analysis of the turn symmetry statistics for the artificial walker. (A) Gaussian model used to generate the artificial trajectory. (B) Analysis of the turn symmetry statistics performed on the generated trajectory. (C) Analysis of the turn symmetry statistics based on ten tracks of the artificial walker describing the same generated trajectory.}
        \label{traycompsym}
    \end{center}
\end{figure}

\begin{figure}[!ht]
    \begin{center}
        \includegraphics[width=240px]{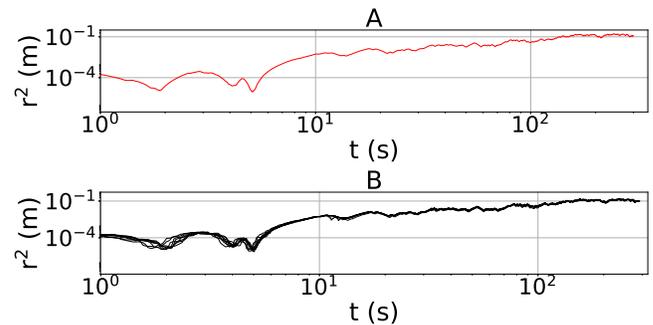}
        \caption{Analysis of diffusivity in the trajectories of artificial walkers. (A) Analysis of the diffusivity performed on the generated trajectory. (B) Analysis of the diffusivity based on ten tracks of the artificial walker describing the same generated trajectory.}
        \label{traycompdiff}
    \end{center}
\end{figure}

There are some qualitative differences in the trajectories generated and estimated. To quantify the accuracy, we use the variables of interest which are computed and shown in the Figures \ref{traycompsym} and \ref{traycompdiff}. 

Figure \ref{traycompsym} shows the statistics of turn symmetry computed after tracking 10 trajectories of the generated walker. It is possible to see how consistent the measurements are. 

Figure \ref{traycompdiff} shows the diffusivity analysis based on the same 10 trajectories, also demonstrating the good repeatability of the measurements.

The results shown suggest that our instrument is performing well in several trackings of an artificial walker. Our data show that even with accumulative errors due to the relative localization systems, the variables of interest (turn symmetry and diffusivity) are not significantly affected except for the addition of some noise, as expected.


\subsection{Tracking of real walkers}
\label{sub_restrackreal}

\begin{figure}[!ht]
    \begin{center}
        \includegraphics[width=240px]{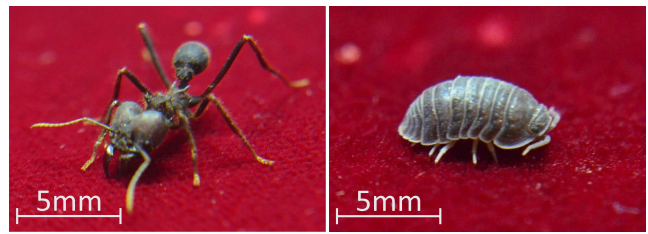}
        \caption{Arthropods used in the experimentation. left: a typical worker of the ant {\it Atta insularis}. Right: isopod \emph{Armadillidium} sp.}
        \label{animals}
    \end{center}
\end{figure}

Several tracking experiments on real arthropods in non-confined regions were performed using our instrument. Here we show the study of two different species of arthropods: the ant \emph{Atta insularis} and the crustacean \emph{Armadillidium} sp., both treatable with the instrument in terms of size. Both species are shown in Figure \ref{animals}. The non-confined regions covered an area of at least three orders of magnitude greater than those of the species' bodies. Both studies were carried out in a quasi-controlled environment. The dimensions of the surface are 10\,m long by 10\,m wide, totaling up to 100\,m$^2$. Those of the arthropods are described in Sections \ref{armadillum} and \ref{atta} and are in the order of 1\,cm. The area is considered non-confined. The surface over which the arthropod forages is free of obstacles and considered plane in its majority. This is a key detail as irregularities in the ground are big obstacles to small animals. The floor has features that can be easily recognized, which are needed to estimate the position of the mobile robotic platform, and therefore the camera, using the MVO method described in \ref{sub:trajrec}.

\subsection{Atta insularis}
\label{atta}

The ant \emph{Atta insularis}, commonly known as \emph{Bibijagua}, is a Cuban endemic ant species \cite{alfredo2016}. This insect has been used in several quantitative studies \cite{nicolis2013foraging,altshuler2005symmetry,tejera2016uninformed,reyes2016preliminary,reyes2016flux}. Their movement has been characterized as super-diffusive when foraging in confined areas\cite{alfredo2016}. 

\begin{figure}[!h]
    \begin{center}
        \includegraphics[width=260px]{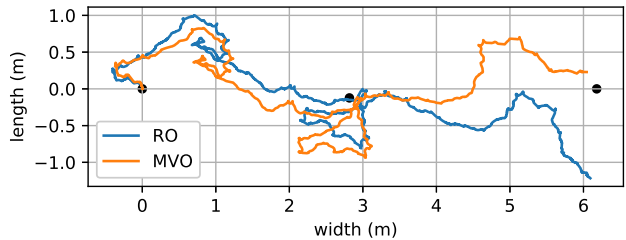}
        \caption{Trajectory of the ant \emph{Atta insularis} relative to the initial position of the insect using the RO and MVO methods of camera tracking. The three black dots represent the actual position of the insect at the beginning, half and end point of the trajectory, respectively, measured with external instruments.}
        \label{fig-chart-atta-insularis}
    \end{center}
\end{figure}

To provide some preliminary data gathered with our experimental system, an ant \emph{Atta insularis} was tracked for 10\,min. During this interval, the insect, as well as the robot, moved 10\,m approximately. Figure \ref{fig-chart-atta-insularis} shows the position of the ant, estimated with both RO and MVO. There is an increasing difference between the trajectories estimated by both methods due to the cumulative errors introduced.

\begin{figure}[!h]
    \begin{center}
        \includegraphics[width=250px]{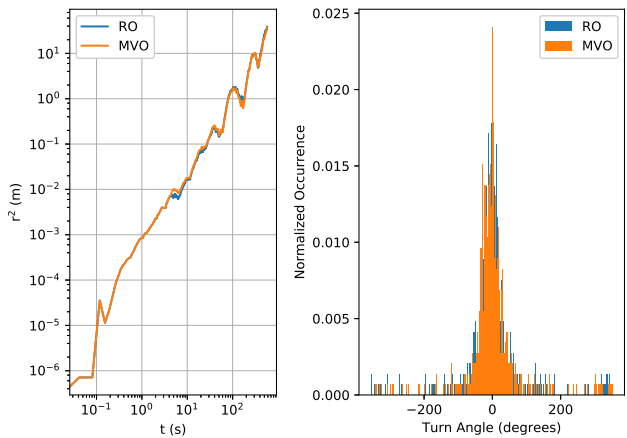}
        \caption{Analysis of (A) diffusivity and (B) Turn Symmetry for the ant \emph{Atta insularis}.}
        \label{fig-chart-atta-insularis-variables}
    \end{center}
\end{figure}

The trajectory was analyzed using the variables mentioned in the Section \ref{sec_measuerd_variables}. The preliminary results obtained are shown in Figure \ref{fig-chart-atta-insularis-variables}. It is possible to corroborate the same regime of diffusivity found by A. Reyes \emph{et al.} for the case of confined regions\cite{reyes2016preliminary}. As expected, The MVO method provides a more accurate trajectory, given the difference with the three reference points represented in Figure \ref{fig-chart-atta-insularis}. It is also possible to check that, even with the differences in the integrated trajectories due to the accumulative errors present in both methods, there are no appreciable differences between the variables computed using both trajectories, which confirms once again that the local features of the trajectories is key for the analysis.

\subsection{Armadillidium sp.}
\label{armadillum}

\begin{figure}[!h]
    \begin{center}
        \includegraphics[width=250px]{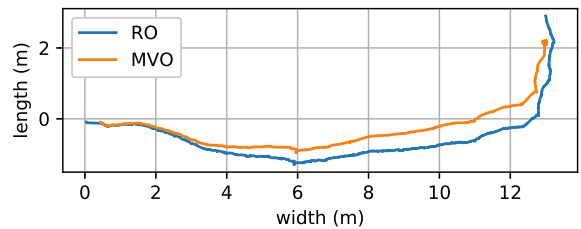}
        \caption{Trajectory of the isopod \emph{Armadillidium} sp. relative to the initial position of the insect using the RO and MVO methods of camera tracking.}
        \label{fig-chart-armadillidium-vulgare}
    \end{center}
\end{figure}

Another track was performed on the isopod \emph{Armadillidium} sp. Similar walkers have been studied in a 0.6\,m$\times$0.6\,m region by others\cite{tuck2004foraging,hassall2005effects}. Our experiment took 11.25\,min. During this interval the walker moved 19.70\,m approximately. Figure \ref{fig-chart-armadillidium-vulgare} shows the estimated trajectory of the isopod also using both methods explained in Section \ref{sub:trajrec}.

\begin{figure}[!h]
    \begin{center}
        \includegraphics[width=250px]{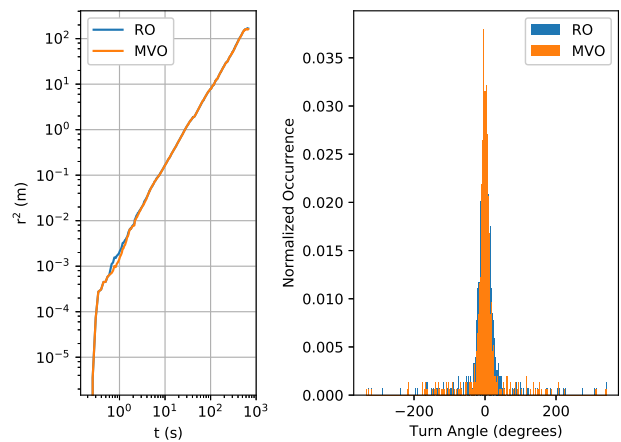}
        \caption{Analysis of (A) diffusivity and (B) Turn Symmetry for an individual of the species \emph{Armadillidium} sp.}
        \label{fig-chart-armadillidium-vulgare-variables}
    \end{center}
\end{figure}

The variables presented in Section \ref{sec_measuerd_variables} were also computed for the case of the isopod \emph{Armadillidium} sp. and the results are shown in Figure \ref{fig-chart-armadillidium-vulgare-variables}.

\section{Conclusions}
\label{conclu}

The main contribution in this paper is the design, construction and validation of an original instrument for tracking millimetric-sized walkers, which is capable of collecting trajectories of different biological species, and track them over $10$\,m or more, even in darkness. Previous tracking systems typically used a fixed camera, and the walkers were followed within areas no larger than approximately one square meter and other mobile tracking devices had poor spatial resolution\cite{tuck2004foraging,alfredo2016,reyes2016preliminary,DGPS2013}.
 
We have shown the effectiveness of our robot in two ways. Firstly, by repeatedly tracking artificial walkers on a LCD screen, acheiving good repeatibility of parameters of biological interest. Secondly, by performing a preliminary tracking of two arthropod species: workers of the leaf-cutter ant {\it Atta insularis} and individuals of the species \emph{Armadillidum} sp. The analysis of diffusivity and turn symmetry statistics based on the obtained tracks is consistent with previous data obtained in smaller areas in the case of {\it Atta insularis} (analogous data for the \emph{Armadillidum} sp. is not available in the literature, to our knowledge).
 
We believe that, by systematically applying the new tracking tool to a range of animal species, new light can be shed on long-standing biological puzzles, such as determining the precise mechanisms of orientation of millimetric-sized walkers, especially arthropods.

\section{Acknowledgements}
\label{acknowledgements}
The authors appreciate insightful discussions and data provided by M. Curbelo and A. Haidar. This work was partially supported by the University of Havana's project ``Active matter: quantification of individual and collective dynamics''.


\end{document}